\documentclass[10pt,letterpaper]{article}
\usepackage{opex3}
\usepackage{cite}
\usepackage{amsmath}
\usepackage{graphicx}
\usepackage{amsfonts}
\usepackage{amssymb}
\usepackage{dcolumn}
\usepackage{amsmath}
\usepackage{color}

\begin{document}
\title{A perfect absorber made of a  graphene micro-ribbon metamaterial}

\author{Rasoul Alaee,$^*$ Mohamed Farhat, Carsten Rockstuhl, and Falk Lederer}

\address{Institute of Condensed Matter Theory and Solid State Optics,
Abbe Center of Photonics, Friedrich-Schiller-Universit\"{a}t
Jena, Max-Wien-Platz 1, D-07743 Jena, Germany}
\email{$^*$Rasoul.Alaee@uni-jena.de}

\begin{abstract} Metamaterial-based perfect absorbers promise many applications.
Perfect absorption is characterized by the complete suppression
of transmission and reflection and complete dissipation of the
incident energy by the absorptive meta-atoms. A certain
absorption spectrum is usually assigned to a bulk medium and
serves as a signature of the respective material. Here we show
how to use graphene flakes as building blocks for perfect
absorbers. Then, an absorbing meta-atom only consists of a
molecular monolayer placed at an appropriate distance from a
metallic ground plate. We show that the functionality of such
device is intuitively and correctly explained by a Fabry-Perot
model.
\end{abstract}

\ocis{(050.6624) Subwavelength structures;(160.3918) Metamaterials;
(310.3915) Metallic, opaque, and absorbing coatings.}


\section{Introduction}
Graphene is a two-dimensional (2D) carbon material with
tremendous applications. In 1962 it was isolated by Boehm
\textit{et al.} for the first time \cite{GR0}. More recently,
in 2004, a team led by Geim proposed a more efficient way to
produce it based on exfoliation procedures. That moment was the
starting point for the rise of graphene \cite{GR1}. The
research was fueled by the extraordinary electronic transport
properties of graphene which differ substantially from those in
metals and semiconductors \cite{GR2}. In fact, electrons behave
like massless particles when they propagate in graphene sheets,
like photons do in matter with the notable difference that they
have an electric charge. These unprecedented electronic
properties make graphene a serious candidate for the material
of the $\text{21}^{\text{st}}$ century \cite{GR3, GR4} to be
used for ultrafast, low loss electronic devices because of its
high conductivity and its extraordinary mechanical parameters.

Likewise the optical properties of graphene attracted a great
deal of research interest. In the visible, graphene acts as an
absorptive dielectric. The absorption \emph{A} of a single
sheet amounts to $A\approx\pi\alpha\approx2.3\%$
\cite{FSTR,GR2}, with $\alpha=e^2/\hbar c\approx1/137$ being
the fine structure constant. Various applications, ranging from
graphene based optical modulators \cite{GRP10}, transformation
optics \cite{GRP9} to numerous other devices
\cite{GRP_chen,GRP11,GRP12} may be envisaged by using graphene
in the optical domain. Examples have been already
experimentally demonstrated and have been provided solid
evidence for the versatility of graphene
\cite{GRP7,GRExp1,GRExp2}.

At lower frequencies, the optical properties of graphene
resemble those of a Drude-type material. Such materials promote
the formation of surface plasmon polaritons (SPP). An SPP is a
collective oscillation of the charge density and light at the
interface between a graphene sheet and its surrounding
\cite{GRExp1,GRExp2,GRP1,GRP2,GRP4,GRP8}. An SPP supported at a
graphene interface can be guided in many ways like other types
of waves (light or sound). This property attracted a particular
interest of the metamaterial and plasmonic community
\cite{GRP5,GRP13}.

One of the most interesting properties of plasmonic
metamaterials is the perfect absorption of light \cite{MPA1,
MPA_REV}. Such functionality was demonstrated while relying on
different designs and while optimizing the point of operation
over large frequency ranges
\cite{Abajo,MPA3,BMPA3,MPATE,MPAPS,BMPA1,Shvets}. The
functionality of such perfect absorbers can be understood in
the framework of the more general theory of coherent perfect
absorption (CPA). It is in many ways the time-reversed analogue
of the lasing effect. Instead of radiating a coherent flux of
light in a very narrow spectral band \cite{CPA1,CPA2}, CPA can
occur for structures possessing intrinsic losses. It only
requires that an incoming radiation pattern is dissipated by
suppressing the associated reflection and transmission
channels. In fact, the CPA effect can only occur because of the
interplay of interference and absorption being achievable just
for a very specific geometry and frequency range. More
recently, graphene nanodisks were suggested for this purpose
\cite{MPAGR}. It was shown that graphene flakes allow for the
perfect absorption of THz radiation owing to their high
extinction cross section (compared to their geometrical cross
section). Such perfect absorption has also been achieved when
the discs were put in front of a mirror at some distance.

Recently graphene micro-ribbons is theoretically and
experimentally studied \cite{GRVidal, GRP7}. In this paper we
propose to take advantage of the recent technological progress
in fabricating nanostructured graphene layers grown on
dielectric substrates \cite{GRP7}. We show that perfect
absorption can be achieved by patterned graphene micro-ribbons
on a thick dielectric layer deposited on top of a reflecting
metal substrate. Furthermore, we show that the complete light
absorption is almost independent of angle of incidence and its
frequency can be easily tuned by changing the chemical
potential or gate voltage.

The \textit{modus operandi} of the suggested perfect absorber
differs from that of the ordinary ones made of metallic
nanoparticles brought in close proximity to a metallic ground
plate. There, the current induced in the ground plate via a
near-field interaction is out-of-phase to the current in the
nanoparticle. This causes the resonance to be dark and
radiative losses to be reduced. If the condition of critical
coupling is achieved, the incident light can be fully absorbed
\cite{MPA5}. In striking contrast, the device we suggest and
study here is characterized by a thick dielectric intermediate
layer that prohibits any near-field interaction between the
nanostructured graphene and the ground plate.  This type of
perfect absorber is known as the Salisbury screen or absorber
\cite{Salisbury}. Instead, perfect absorption is achieved by
exploiting a perfect destructive interference of the reflected
light \cite{MPA_REV,Salisbury, MPA_DestInter1, MPA_DestInter2}.
Transmission through the structure is totally suppressed
because the thickness of the ground plate is much larger than
the typical skin depth at THz frequency. Therefore, the
complete electromagnetic energy gets absorbed by the graphene
micro-ribbons. We explain our findings by considering the
micro-ribbons as a metasurface. This allowed us to describe
them only by their reflection and transmission coefficients and
using Airy formula for an asymmetric Fabry Perot cavity to
calculate the absorption in the device \cite{Airy}. Predictions
from such simple model are in excellent agreement with
full-wave simulations and clearly prove that the perfect
absorption is a purely coherent effect.

\section{Graphene perfect absorber}
\begin{figure}[h]
\centering
\includegraphics[width=95mm,angle=0] {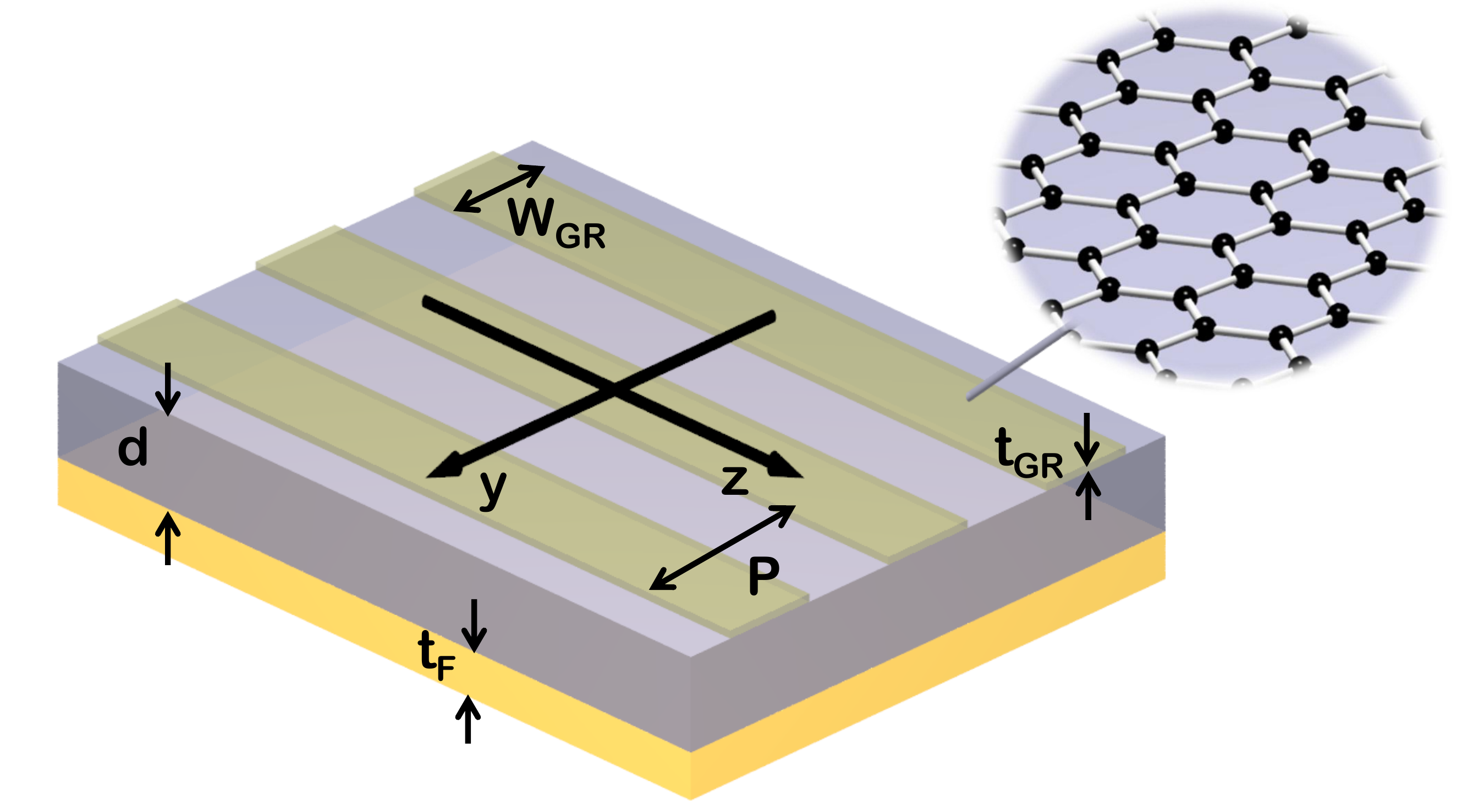}
\caption[Submanifold]{Schematic of the graphene
micro-ribbon perfect absorber. The geometrical parameters of the
proposed structure are: $t_{\mathrm{F}}=1\,$ $\mu$m,
$d=1-100\,$ $\mu$m, $W_{\mathrm{GR}}=1\,$ $\mu$m and
$P=2$ $\mu$m. The inset shows the honeycomb microscopic structure of
the graphene layer arranged as micro-ribbons.} \label{FIG_GPA}
\end{figure}

The structure under consideration is shown in Fig.
\ref{FIG_GPA}. It consists of a graphene micro-ribbon array on
top of a metallic ground plate separated by a thick dielectric
spacer. It is periodic in one direction ($y$) with periodicity
$P=2 \mu$m and infinitely extended in the other one. We assume
the refractive index ($n=\sqrt{\varepsilon_{\mathrm{d}}}=2.1$)
of the dielectric deposited on the metal. The ground plate is
made of gold with a conductivity $\sigma=4\times10^{7}$S/m
which is perfectly reflecting in the frequency domain of
interest (Far-infrared regime).

\begin{figure}[h]
\centering
\includegraphics[width=95mm,angle=0] {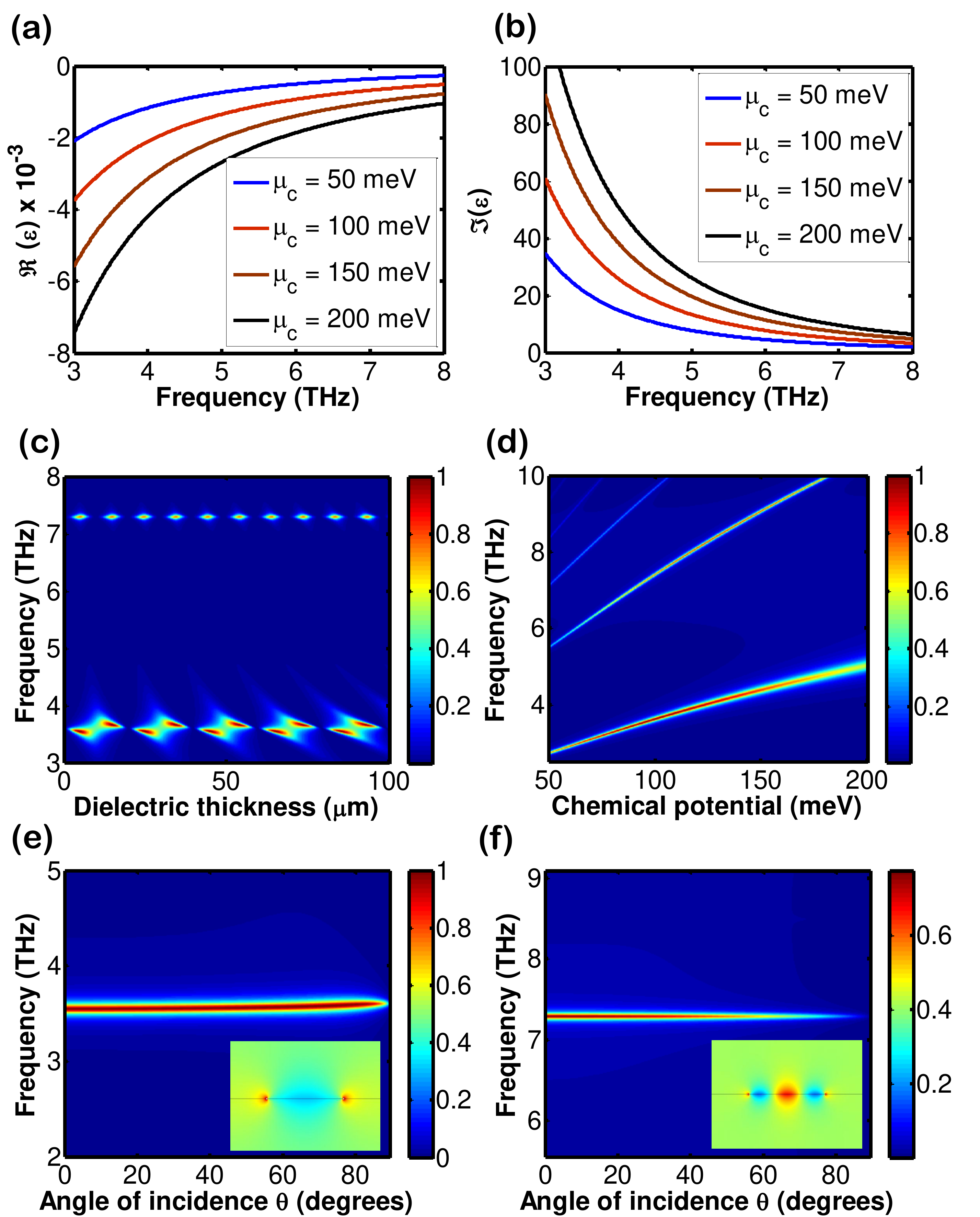}
\caption[Submanifold]{(a) Real and (b) imaginary parts of the relative
permittivity of a 1nm thick graphene sheet as function of frequency as well as chemical
potential. (c) Absorption as a function of frequency ($\omega$) and dielectric thickness ($d$) for
a graphene micro-ribbon perfect absorber with chemical potential
$\mu_{\mathrm{c}}=100\,$meV at normal incidence. (d) Absorption as a
function of frequency and chemical potential for graphene
micro-ribbons perfect absorber with $d=4.7\,$ $\mu$m at
normal incidence. (e) and (f) show the absorption as a function of the
angle of incidence with dielectric thickness $d=4.7\,$
$\mu$m and $d=4.9\,$ $\mu$m for the first and second
mode, respectively. The insets show the electric field distribution
$E_y$ for both modes as well.} \label{FIG_AbsGPA}
\end{figure}

The graphene was numerically modeled by a thin layer ( with
thickness $\Delta=1$nm) of permittivity
$\varepsilon_{\textrm{GR}}=\varepsilon_0+i\sigma_{\textrm{GR}}/(\omega\Delta)$,
with $\sigma_{\textrm{GR}}$ as the surface conductivity of the
graphene sheet \cite{GRP9} which can be derived using the
well-known Kubo formula \cite{kubo1,kubo2,kubo3,kubo4} and
writes as:

\begin{eqnarray}
\sigma_{\textrm{GR}}&=&\frac{ie^2}{4\pi\hbar}\ln\left[\frac{2|\mu_c|-(\omega+i2\Gamma)\hbar}{2|\mu_c|+(\omega+i2\Gamma)\hbar}\right]
+\frac{ie^2k_\mathrm{B}T}{\pi\hbar^2(\omega+i2\Gamma)}\left[\frac{\mu_c}{k_\mathrm{B}T}+2\ln(e^{-\mu_c/k_\mathrm{B}T}+1)\right],
\label{kubo}
\end{eqnarray}

\noindent where $e$, $\hbar$ and $k_\mathrm{B}$ are universal
constants related to the electron charge, Planck's and
Boltzmann's constant, respectively. $T$ is the temperature and
is fixed to 300 K. $\mu_c$ and $2\Gamma$ ($2\Gamma=\hbar/\tau$,
$\tau$ is the electron-phonon relaxation time) are physical
parameters of the graphene sheet and account for the chemical
potential (or Fermi energy) and the intrinsic losses,
respectively. We assumed $\Gamma = 0.1$ meV in this
contribution and this value is based on the theoretical
estimation of maximum mobility in graphene \cite{kubo1}.
Furthermore, it should be noted that even for a larger
scattering rate the general functionality of proposed device is
not affected and all the phenomena can be observed. The
frequency of operation $\omega$ is in the Far-infrared regime
(1-10 THz). The parameter $\Delta$ was taken 1 nm in our study,
and the convergence of $\varepsilon_{\textrm{GR}}$ was verified
numerically (by taking the limit $\Delta\rightarrow 0$) and by
reproducing some previous results \cite{GRP5,GRP7,MPAGR}. In
addition, the frequency dependent permittivity of graphene
depending on the some selected values of the chemical potential
are shown in Fig. \ref{FIG_AbsGPA}(a) and \ref{FIG_AbsGPA}(b).

The geometrical parameters of our structure are selected based
on experimental capabilities \cite{GRP7}, a filling fraction
$W_{\textrm{GR}}/P=0.5$ was shown to be easier for fabrication
(see Fig. \ref{FIG_GPA}). Nevertheless, the very parameters are
of marginal importance and variations would only cause a slight
modification of the frequency of operation. The set-up is
illuminated by a transverse magnetic (TM) incident plane wave.
The Fourier Modal Method (FMM) is used to explore the
underlying physics of the absorber \cite{FMM}. The method
solves Maxwell's equations for the periodic structure
rigorously. In order to have convergent results, 101 Fourier
orders were selected in all numerical results presented herein.

Figure \ref{FIG_AbsGPA}(c) shows the absorption \emph{A} of the
graphene micro-ribbons (of Fig. \ref{FIG_GPA}) as function of
the thickness of the spacer $d$ and the frequency of operation.
The strong coupling of the Fabry-Perot resonance with the
localized eigenmode of the graphene micro-ribbons is clearly
visible in Fig. \ref{FIG_AbsGPA}(c) and this leads to an
avoided crossing and a significant Rabbi splitting. The results
reveal that perfect absorption ($A=1$) can be achieved at
various values of $d$, periodically spaced between 1 and 100
$\mu$m around a frequency of 3.6 THz for the first mode
(dipolar one) and around 7.3 THz for the second mode, where the
peaks are less pronounced ($A<1$). However, by optimizing the
geometry and tuning the chemical potential, we have verified
that it is possible to achieve total absorption for both modes
(this is not included in this contribution). Moreover, the
resonance frequency of maximum absorption is slightly changing
for different dielectric spacer and this effect is more
pronounced for the first mode (see Fig. \ref{FIG_AbsGPA}(c)).
In Fig. \ref{FIG_AbsGPA}(d) the absorption as a function of the
chemical potential $\mu_c$ (ranging from 50 to 200 meV) and the
frequency is displayed. It can be clearly seen that the
resonance frequency of the absorber can be tuned over a wide
range by varying $\mu_c$. It should be mentioned that by
applying a gate voltage (a static electric field) or by means
of chemical doping, the chemical potential and thus the
conductivity of graphene can be controlled on purpose.

In order to investigate the robustness of the graphene perfect
absorber, the absorption as a function of frequency and the
angle of incidence is displayed in Fig. \ref{FIG_AbsGPA}(e) and
\ref{FIG_AbsGPA}(f). The first mode is almost unaffected while
varying the angle of incidence $\theta$ until 80 degrees
{[}Fig. \ref{FIG_AbsGPA}(e){]}. Then absorption starts to
decrease. The behavior of the higher order mode (the second
mode for instance) is quite different, as shown in Fig.
\ref{FIG_AbsGPA}(f), where the absorption is strongly dependent
on the angle of incidence and significantly decreases beyond 50
degrees to vanish completely at about 80 degrees. The electric
field distribution $E_y$ (component parallel to the gap) at
plasmon resonance  is displayed for both modes in the insets of
Fig. \ref{FIG_AbsGPA}(e) and \ref{FIG_AbsGPA}(f), respectively,
and confirms the nature of the excited modes [dipole and higher
order].
%
%

\section{Mechanism of complete absorption}
\begin{figure}[h]
\centering
\includegraphics[width=95mm,angle=0] {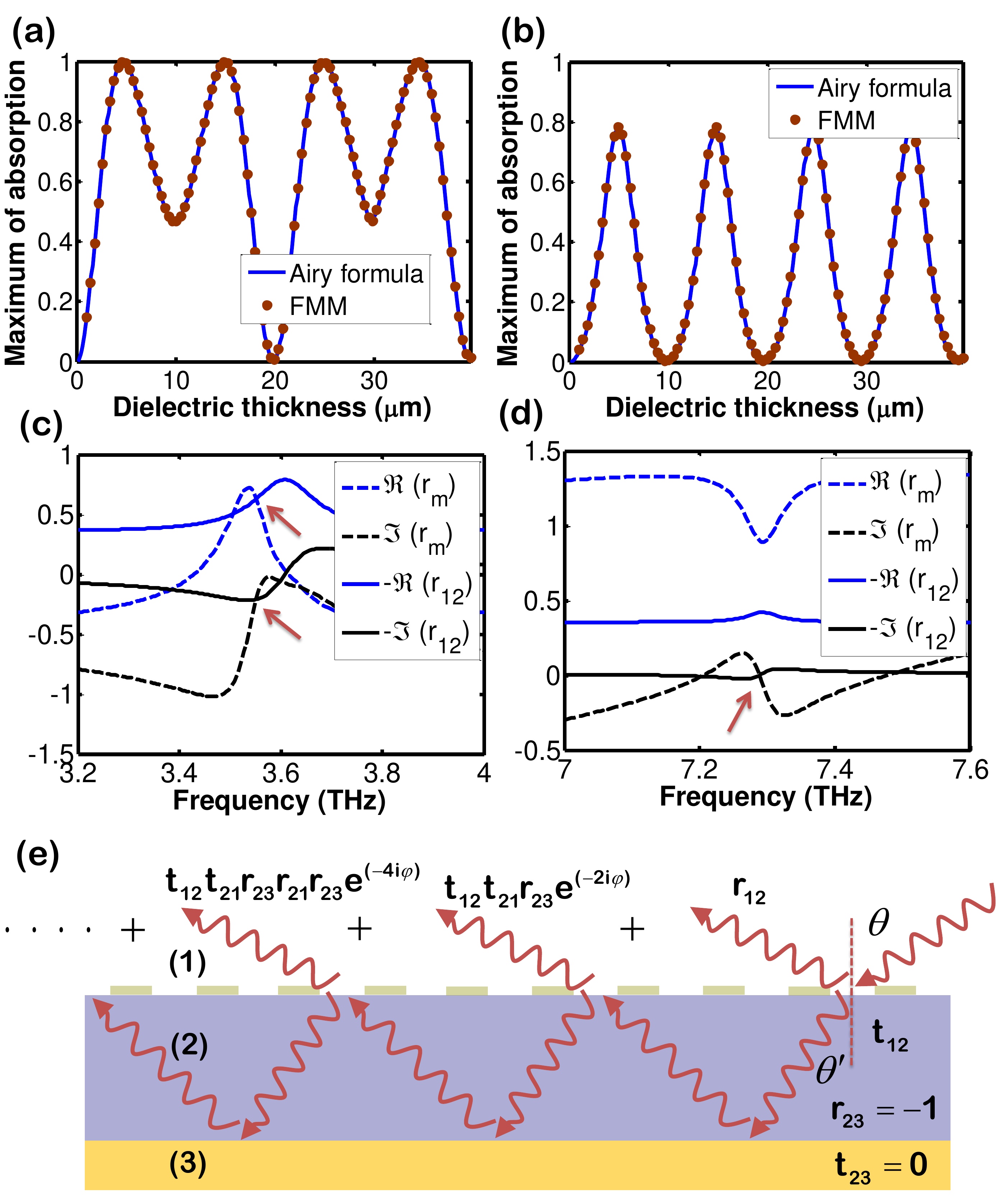}
\caption[Submanifold]{(a) and (b) show maximum
absorption as a function of dielectric thickness and frequency
calculated from semi-analytical approach (Airy formula) and full
wave simulation (FMM) for the first and second mode, respectively.
(c) and (d) show real and imaginary parts of direct reflection
coefficient ($r_{12}$) as well as multiple reflection coefficient
($r_{m}$) with dielectric thickness $d=4.7\,$ $\mu$m and
$d=4.9\,$ $\mu$m for the first and the second mode
respectively. The red arrows represent the crossing point when real
or imaginary parts of reflection coefficients has same magnitude
with opposite sign ($r_{12} = - r_{m}$).(e) Schematic of the
asymmetric Fabry-Perot cavity with the definition of the multiple
reflection and transmission coefficients involved and an indication
on the contribution of the directly reflected light and the
reflected light due to multiple scattering inside the Fabry-Perot
cavity..}\label{FIG_AIRY&FMM}
\end{figure}

Let us focus now on the underlying mechanism of perfect
absorption for graphene. To simplify the structure and to
provide a physical explanation, we shall consider the structure
as an asymmetric Fabry-Perot cavity with two mirrors, i.e., a
graphene micro-ribbon array as the top mirror and the metallic
ground plate as the bottom mirror. The transmission channel
($T$) of the system is completely suppressed by choosing a
sufficiently thick ground plate ($T \simeq 0$). In order to
achieve total absorption of the incident energy ($A=1-T-R
\simeq 1$) the reflection channel must be closed as well ($R =
|r|^2 = |r_{12} + r_{m}|^2 \simeq 0$) which is the sum of the
direct reflection coefficient $r_{12}$ and the multiple
reflection coefficient  $r_{m}$ {[}as sketched in Fig.
\ref{FIG_AIRY&FMM}(e){]}. The total reflection coefficient can
be expressed as \cite{Airy}:
\begin{equation}
r=r_{\mathrm{12}}+r_{\mathrm{m}}=r_{\mathrm{12}}+\frac{t_{\mathrm{12}}
t_{\mathrm{21}}r_{\mathrm{23}}e^{(-2i\varphi)}}{1-r_{\mathrm{21}}r_{\mathrm{23}}e^{(-2i\varphi)}},
\label{EQ_Airy}
\end{equation}
where, $t_{12}$, $t_{21}$, $r_{12}$, $r_{21}$, $r_{23}$ are
complex-valued transmission and reflection coefficients at both
interfaces, $\varphi= k_0nd\cos\theta^{\prime}$ is the phase
accumulated upon a cavity transfer, $k_0$  is the free space
wavenumber, $n$ is the refractive index of the dielectric, and
$d$ is its thickness.
The dependence of the absorption maxima on the dielectric
spacer thickness $d$ calculated by using the Airy formula (Eq.
\ref{EQ_Airy}) is shown in Fig. \ref{FIG_AIRY&FMM} for the
first (a) and the second (b) modes. This formula could be used
by considering the graphene micro-ribbons as a metasurface
(this could be justified by its extremely small thickness
compared to other dimensions and to the wavelength), and it is
completely defined by its reflection and transmission
coefficients. These coefficients were calculated using FMM
while assuming that the sufficient thin graphene metasurface is
sandwiched between two semi-infinite half-spaces; one of them
being air; the other being the dielectric. Assuming an
excitation with a linearly polarized incident plane wave either
from the top medium (air) or from the bottom medium
(dielectric), provides unambiguously the coefficients $r_{12}$,
$t_{12}$, $r_{21}$, and $t_{21}$, respectively.

For comparison, the rigorous calculations (using FMM) are also
displayed in the same figure, and a perfect agreement between
the two methods is evident for both modes [see
Fig.\,\ref{FIG_AIRY&FMM}(a) and \ref{FIG_AIRY&FMM}(b)]. In
addition, the real and imaginary parts of the multiple
reflection coefficient $r_{m}$ and the direct reflection
coefficient $r_{12}$ are plotted in Fig. \ref{FIG_AIRY&FMM}(c)
and \ref{FIG_AIRY&FMM}(d) again for both modes. Let us
concentrate for a while on the first mode {[}Fig.
\ref{FIG_AIRY&FMM}(c){]}; at the resonance frequency $f=3.6$
THz, and for a spacer thickness $d=4.7\mu$m, both coefficients
$r_{m}$ and $r_{12}$ have the same phase and opposite
amplitude. Hence, the resulting total reflection coefficient is
completely canceled due to destructive interference ($r_{12} =
- r_{m}$). As a result, the absorption is almost unity for this
frequency. However, we expect for the second mode, resonant at
$f=7.3\,$ THz that the multiple reflection coefficient can not
totally eliminate the direct reflection coefficient at
$d=4.9\,$$\mu$m because the absorption is less than unity,
{[}see Fig.\,\ref{FIG_AIRY&FMM}(d){]}. But as mentioned
earlier, this issue could be fixed by optimizing the structure
which is beyond the scope of this paper.\\

\section{Conclusion}
To conclude,  in this contribution we have discussed a new
class of perfect absorbers in the Far-infrared regime based on
graphene micro-ribbons. Our structure showed complete
absorption for the first mode and for different geometrical
configurations. The robustness of the mechanism was evidenced
by analyzing its dependence on the angle of incidence and the
chemical potential of the graphene sheets. Moreover, we
utilized a semi-analytical approach based on a Fabry-Perot
model that explains very well the observed behavior. We believe
that our study demonstrates the versatility of graphene and
could be used in future for optical or THz interconnects.\\
\section*{Acknowledgments}
This work was partially supported by the German Federal Ministry of
Education and Research (Metamat and PhoNa) and by the Thuringian
State Government (MeMa).
\end{document}